# A pump-probe XFEL particle injector for hydrated samples

*U. Weierstall, R.B. Doak, J.C.H. Spence*

*Department of Physics, Arizona State University, Tempe, AZ 85287-1504, USA*

**Abstract**

We have developed a liquid jet injector system that can be used for hydrated sample delivery at X-ray Free Electron Laser (XFEL) sources and $3^{rd}$ generation synchrotron sources. The injector is based on the Gas Dynamic Virtual Nozzle (GDVN), which generates a liquid jet with diameter ranging from 300 nm to 20 μm without the clogging problems associated with conventional Rayleigh jets. An improved nozzle design is presented here. A differential pumping system protects the vacuum chamber and an in-vacuum microscope allows observation of the liquid jet for diagnostics while it is being exposed to the X-ray beam. A fiber optically coupled pump laser illuminating the jet is incorporated for pump-probe experiments. First results with this injector system have been obtained at the LCLS.

**Introduction**

Recently X-ray Free Electron Laser (XFEL) sources have become available for user experiments ranging from atomic physics to biological structure determination. Compared to $3^{rd}$ generation synchrotron sources FEL's have about 10 orders of magnitude higher peak brilliance and the photons are concentrated into pulses of femtosecond duration and micrometer size. The first VUV and soft X-ray FEL to come online in 2005 was the FLASH facility at DESY in Germany. Currently it covers a wavelength range from 4.5 nm to about 47 nm with GW peak power and pulse durations between 10 fs and 100 fs. In 2009 the Linac Coherent Light Source (LCLS) came online and was providing X-rays to users with 0.7 nm wavelength and pulse lengths of 3 to 300 fsec [1]. It has been proposed [2], that biomolecules can be imaged with an FEL beyond the traditional radiation damage limits if only the X-ray pulse length is short enough to outrun damage effects. The first experimental test examining this proposal has been performed at FLASH in 2006 [3] and recently we have recorded femtosecond diffraction patterns of biological material at the LCLS [4, 5]. The first particle injector used at FLASH was a modified aerosol injector developed at LLNL [6, 7]. It allows injection of dehydrated or partially hydrated particles. This injector uses an aero dynamic lens to focus aerosol particles at the X-ray interaction region, where the particle beam focus is about 200um wide. Due to the low concentration of particles in the interaction area, the hit rate, i.e. the number of detected diffraction patterns per second divided by the number of X-ray pulses per second is of the order of $10^{-4}$. The advantage of this type of injector is that water surrounding the particle can be completely removed. For objects that require complete hydration, we have developed another injector based on a liquid jet injected into vacuum.

# Design of the injector

## a. Nozzle design

The injector design makes use of the previously described Gas Dynamic Virtual Nozzle (GDVN) [8-10], which produces a liquid jet of micron to submicron diameter without the clogging problems associated with a Rayleigh jet of the same diameter. This nozzle uses a co-flowing sheath gas to reduce the diameter of the liquid jet. As the jet emerges from a tapered 50 micron ID capillary, it is focused down by the sheath gas to a few micron diameter (Figure 1). The jet then breaks up into droplets.

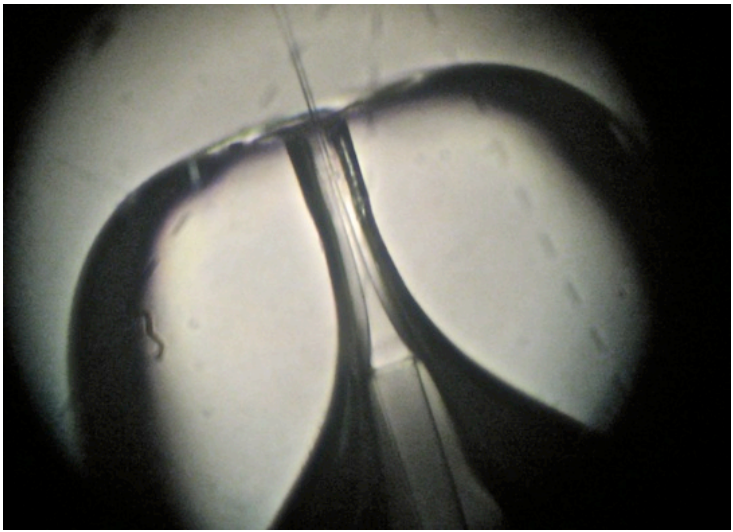

Figure 1:
Gas Dynamic virtual nozzle. A capillary with 50 micrometer ID is centered inside a glass tube, which is flame polished at the end to form a 100 micrometer diameter aperture. Gas flows parallel to the liquid jet and accelerates the liquid, thereby reducing the diameter of the liquid jet.

This type of nozzle has been used during the last few years for experiments at the Advanced Light Source (ALS) at Lawrence Berkeley National Laboratory (LBNL). Protein powder diffraction data has been collected in preparation for the LCLS experiments from Photosystem I nanocrystals injected into the soft X-ray beam at the ALS [11] [12].

We have made some improvements to the miniature GDV nozzles, which will be described here. The previous miniature nozzle design (Figure 1) had the disadvantage, that the centering of the inner capillary in the outer glass tube could not be reproducibly achieved. The new design uses an outer glass tube with a square cross section for the gas flow, which has an inner wall to wall distance of 400 micrometer. The end of this tube is flame polished to obtain a round gas exit aperture with an ID of about 100 micron diameter. After flame polishing, the inner surface leading to the exit hole tapers continuously from a square cross section to a round shape, i.e. it is shaped like a pyramid, which slowly loses its sharp edges (Figure 2). The liquid flow capillary has an OD of 360 micrometer (Polymicro Fused Silica tubing, ID = 50 micron), i.e. it can only move by 40 micrometer inside the square glass tube. This already restricts the possible misalignment between the exit hole and the liquid capillary. Further centering is possible by shaping the liquid capillary end. The liquid capillary end is ground into a cone with a 15 degree taper angle. This is done by slowly rotating the capillary with a motor along its axis while touching the end with rotating abrasive paper under microscope control on a polishing machine (Allied High Tech Techprep 8). After the capillary is inserted into the outer glass tube, the capillary cone surface touches the pyramid sides of the flame polished glass tube and is thereby centered relative to the exit hole. The gas can still flow through the gaps between the pyramid edges and the round liquid nozzle cone. This arrangement leads to a very symmetric liquid exit area. Despite that symmetry, the liquid jet does not necessarily emerge straight, i.e. on the nozzle axis. In Figure 3, two frames out of a movie are show visualizing the emerging liquid jet while all external parameters are held constant. The end of this nozzle has been ground off after the liquid capillary has been installed and glued in place, which leads to a very symmetric nozzle exit area. Nevertheless abrupt changes in flow direction could be observed while the attachment point of the liquid at the capillary changes from one side to the other. It seems that when the capillary end is flush with the gas exit aperture [9], the point where the liquid attaches to the capillary has to be well centered in the outer glass aperture to obtain a straight jet. The location of this attachment point can change randomly and can lie anywhere on the circumference of the capillary exit hole.

Therefore another liquid nozzle shape has been investigated which provides a unique attachment point that is centered in the gas aperture. This new design is shown in Figure 4. The bore of the capillary is curved and exits on the side of the polished cone. The cone has a sharp tip, which serves as the attachment point for the liquid, i.e. the liquid exits on the side of the cone and is driven to the sharp pointed end by the sheath gas pressure forces. With this capillary end shape there is only one unique attachment point for the liquid, and this point is centered inside the gas exit hole as the cone touches the pyramidal shape of the gas aperture (Figure 5). Liquid jetting from such a nozzle in vacuum is shown in Figure 6. The liquid cone emerging from the tip of the liquid capillary cone is visible at the exit of the nozzle. In vacuum, these nozzles operate even when the liquid cone tip is outside the gas aperture plane (i.e. in the free jet expansion area) as shown in Figure 7, although the diameter of the produced jet is larger. The flush nozzle [9] and the protruding cone design have the advantage, that the full length of the unbroken jet is available for experiments. A long unbroken jet is necessary for pump probe experiments where the jet is illuminated by a larger pump laser focus and then probed at a downstream location with the finely focused X-ray probe beam (for more detail see below). When the GDVN is intended for operation in air, the liquid capillary has to be retracted somewhat from the exit aperture plane to provide for adequate gas focusing (shear and radial) forces on the liquid. The optimum liquid capillary position inside the glass tube is determined and adjusted for each nozzle in a test chamber under microscope control in vacuum (or at ambient pressure) while liquid is jetting. Then the liquid capillary is glued in place and sealed inside the outer glass tube. This procedure combined with the use of flame polished square ID glass tubes leads to a success rate of 90% during the nozzle fabrication, meaning that all nozzles emit a jet and most nozzles emit a jet with an emission angle of less than 10 degrees relative to the nozzle axis. During a LCLS beamtime in June 2010 one of those nozzles (with a conventional capillary tip in a square glass tube) lasted one week with 12 hours of operation per day using several different buffer liquids containing protein nanocrystals (see below for details). The particle hit rate was measured to be up to 40% depending on the concentration of the nanocrystal particles in solution. Ideally the particle concentration should be adjusted to yield slightly less then one particle per X-ray interaction volume to avoid double hits.

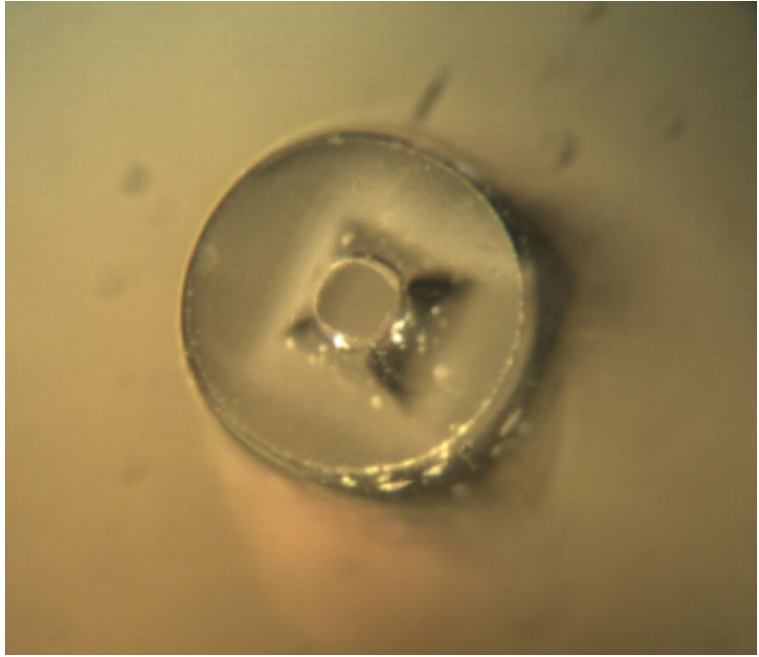

Figure 2: Front view of glass capillary with square ID (and OD) after flame polishing. The rounded end has been ground of to yield a flat exit surface. Below the rounded opening is the pyramidal shape of the inner square glass surface visible. Inner wall to wall distance is 400 micron. Outer wall to wall distance is 0.6 mm.

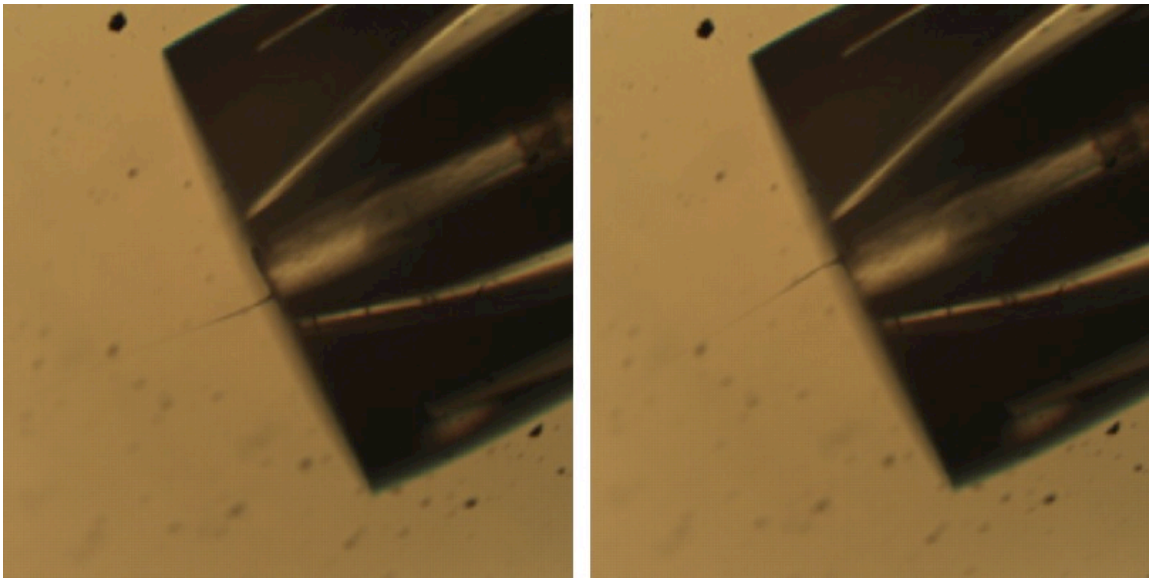

Figure 3: Two frames from a movie showing jetting of a flush nozzle (a nozzle where the liquid capillary and the outer gas capillary end in the same plane). The jet is not straight and moves between two metastable positions whenever the attachment point of the liquid changes.

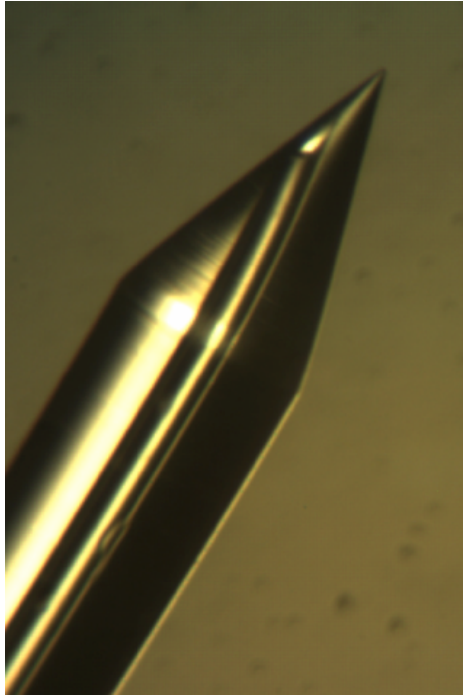

Figure 4: Liquid nozzle shape with asymmetric inner bore. The liquid exits on the side of the cone and is driven to the cone tip by gas forces.

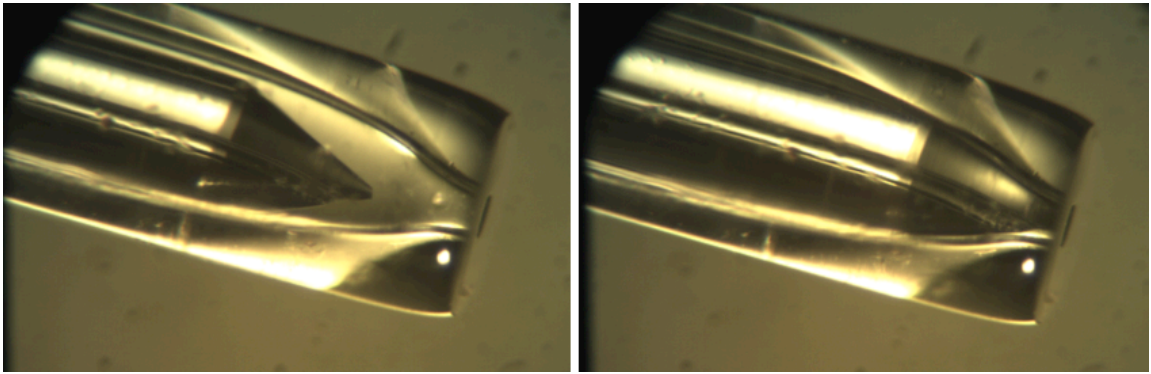

Figure 5: Centering the cone tip with asymmetric bore inside the square ID glass tube. (a) capillary retracted (b) capillary in contact with the pyramid sides at the exit of the glass tube. The pyramid sides center the cone tip inside the exit hole.

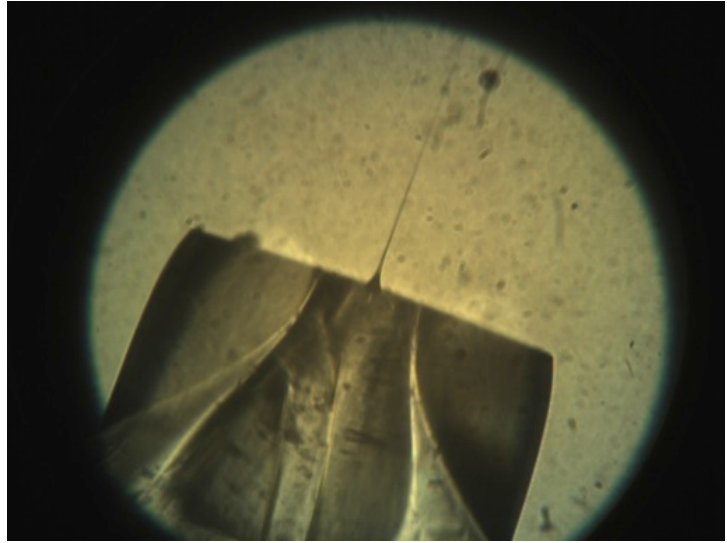

Figure 6: In vacuum jetting of a capillary with asymmetric bore inside a square ID glass tube. Since the cone tip of the liquid capillary is centered inside the gas aperture, the jet emerges straight. The liquid cone attached to the glass cone tip is visible. Square glass tube OD: 0.6mm (wall to wall). Gas exit aperture diameter ~100 micron.

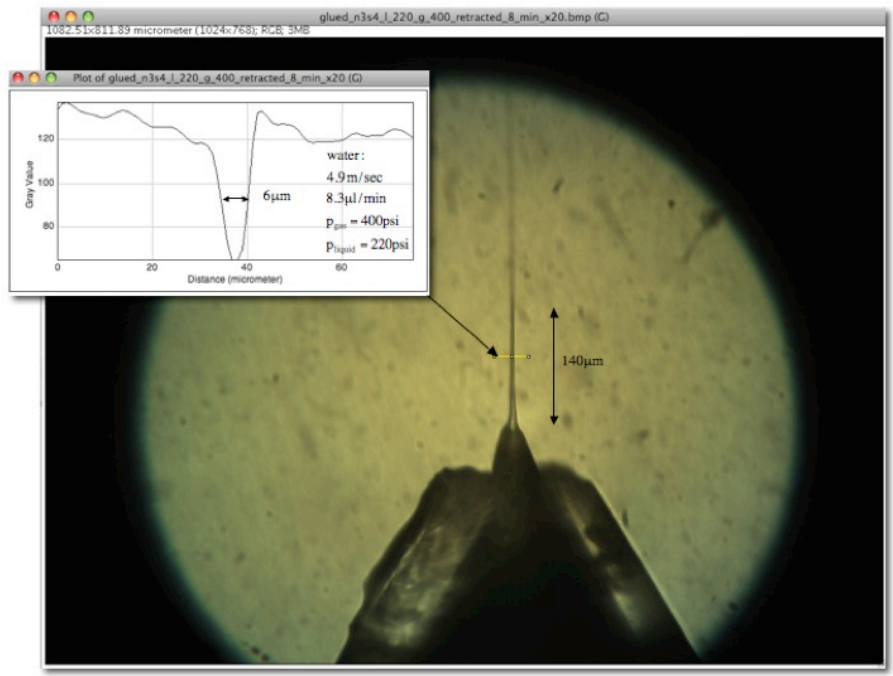

Figure 7: In vacuum jetting of a capillary with asymmetric bore inside a square ID glass tube. The tip of the cone on the liquid capillary protrudes outside the gas aperture into vacuum. The liquid jet starts in the free jet expansion regime of the gas aperture. The outer glass tube has been ground off at an angle to allow unrestricted passage of diffracted X-rays from the point of interaction to the detector CCD. The jet breaks up into droplets where its image turns fuzzy with a breakup length of about 140 micron. The jet diameter has been estimated from a line scan shown as inset to about 6 micron. Liquid flow rate: 8.3 ul/min, jet speed: 4.9m/sec, liquid pressure at supply line: 220 psi, gas pressure at supply line: 400 psi.

**b. Differential pumping system**

The nozzles described in the previous section produce a considerable pressure increase inside a vacuum chamber, since the sheath gas exits together with the liquid at a high flow rate. The vacuum has been measured in a test chamber pumped by a scroll pump. During normal jet operation the vacuum pressure was about $7 \times 10^{-1}$ Torr. If these nozzles are to be used in a high vacuum environment as necessary at a XFEL beamline, differential pumping is necessary to protect the chamber vacuum and upstream components from the high gas flow. Therefore we have designed a differential pumping shroud, which allows recording of diffraction patterns from the interaction of the FEL beam with the jet up to diffraction angles of 60 degrees, while protecting the vacuum of the chamber containing the imaging CCD. The differential pumping system has been designed initially to fit into the CFEL - ASG Multi-Purpose Chamber (CAMP) [13] that has been designed by the Max Planck Advanced Study Group (ASG) within the Center for Free Electron Laser Science (CFEL). But our injector system also fits with minor modifications into the X-ray Pump Probe (XPP) and the Coherent X-ray Imaging (CXI) chambers at the LCLS and has been used at CXI in February 2010.

The differential pumping system consists of a nozzle shroud, which contains the nozzle rod with the GDVN at its end (Figure 8). Figure 9 shows the nozzle shroud alone without the catcher shroud. The nozzle rod slides inside the nozzle shroud on a viton gasket and the rod can be retracted behind a gate valve to allow for nozzle exchange without breaking the chamber vacuum. The nozzle rod is used to move the nozzle into the X-ray interaction region inside the nozzle shroud. The shroud has entrance and exit holes for the X-ray beam with a diameter of 2mm. The exit hole pointing towards the X-ray detector is inside a reentrant exit cone with a cone angle of 60 degrees where the center of the cone is located at the intersection of the X-ray beam and the liquid jet (Figure 10). Therefore the highest diffraction angle, which can exit the shroud is 60 degrees.

The liquid jet is directed into a catcher shroud, which is pumped by a turbo molecular pump. The catcher shroud can be decoupled from the nozzle shroud by means of a bayonet coupling which can be disengaged from outside the vacuum. Once the catcher shroud is coupled to the nozzle shroud, the turbo molecular pump on the catcher side provides the differential pumping vacuum at the nozzle. The nozzle rod is guided and vacuum sealed inside the nozzle shroud by a viton gasket which is fixed at the nozzle rod close to the nozzle ((F) in Figure 11). The nozzle shroud is mounted on the experimental chamber with a 6"-xyz-manipulator (MDC) which allows motorized alignment relative to the X-ray beam with a mechanical resolution of 5 micron. In addition the nozzle can be aligned relative to the nozzle shroud parallel and perpendicular to the X-ray beam by a 2-3/4"-xy-manipulator (MDC) motorized with IMS stepper motors. The resolution of these motors is 1.8 degrees per step. The nozzle manipulator moves 0.025 inch per revolution. The nozzle rod acts as a lever where the viton gasket close to the nozzle serves as the pivot point. A transverse movement at the airside end of the nozzle rod results in a nozzle movement reduced by a lever arm reduction factor of 8.5. Therefore the mechanical resolution for transverse nozzle alignment with the IMS motors is 373 nm. This fine adjustment is necessary since the FEL beam at the LCLS has currently a diameter of 3 micron and the jet diameter is about 4 micron, while the resolution of the external xyz-manipulator (which moves the shroud) is only is 5 micron. Note also that the X-ray beam diameter at the CXI chamber will reach submicron dimensions in the near future.

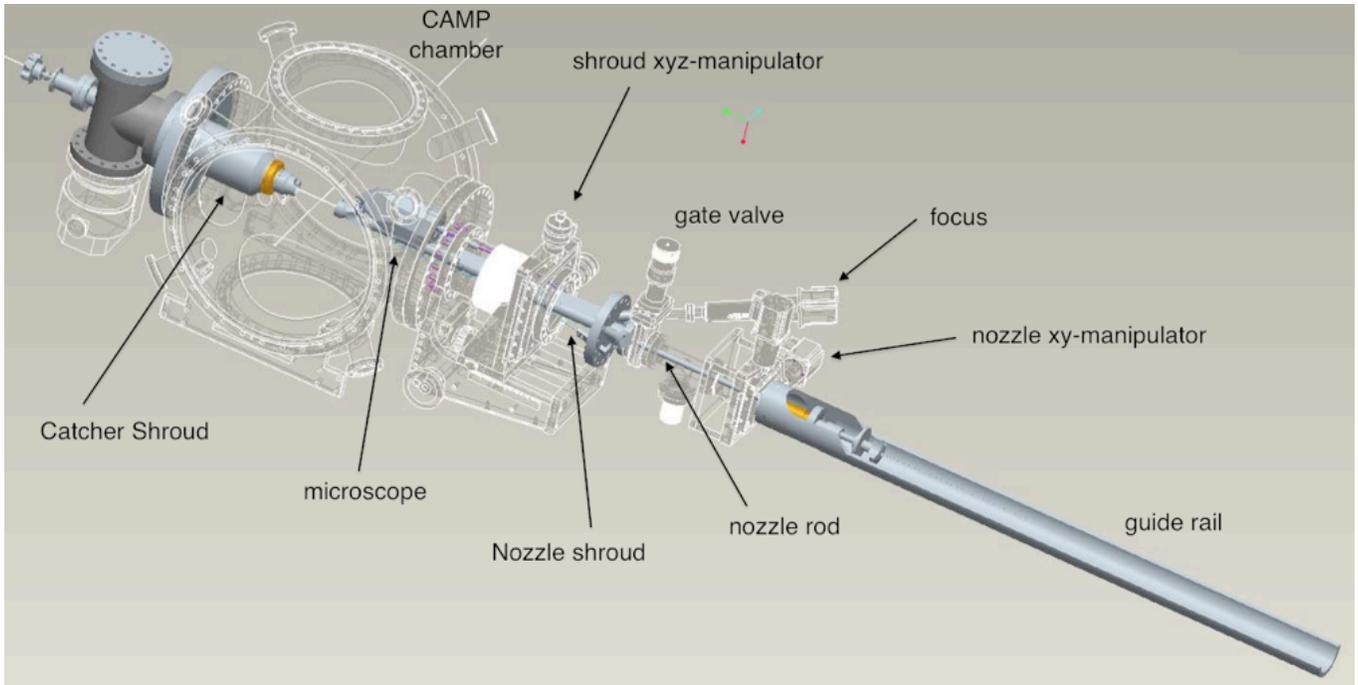

Figure 8: Differential pumping system for the liquid jet injector. It consists of two parts, the catcher shroud and the nozzle shroud, which can be decoupled and retracted via a bayonet coupling. This allows the use of fixed samples without breaking the vacuum. Shown here is the injector inside the CAMP chamber, decoupled at the bayonet and with several parts transparent.

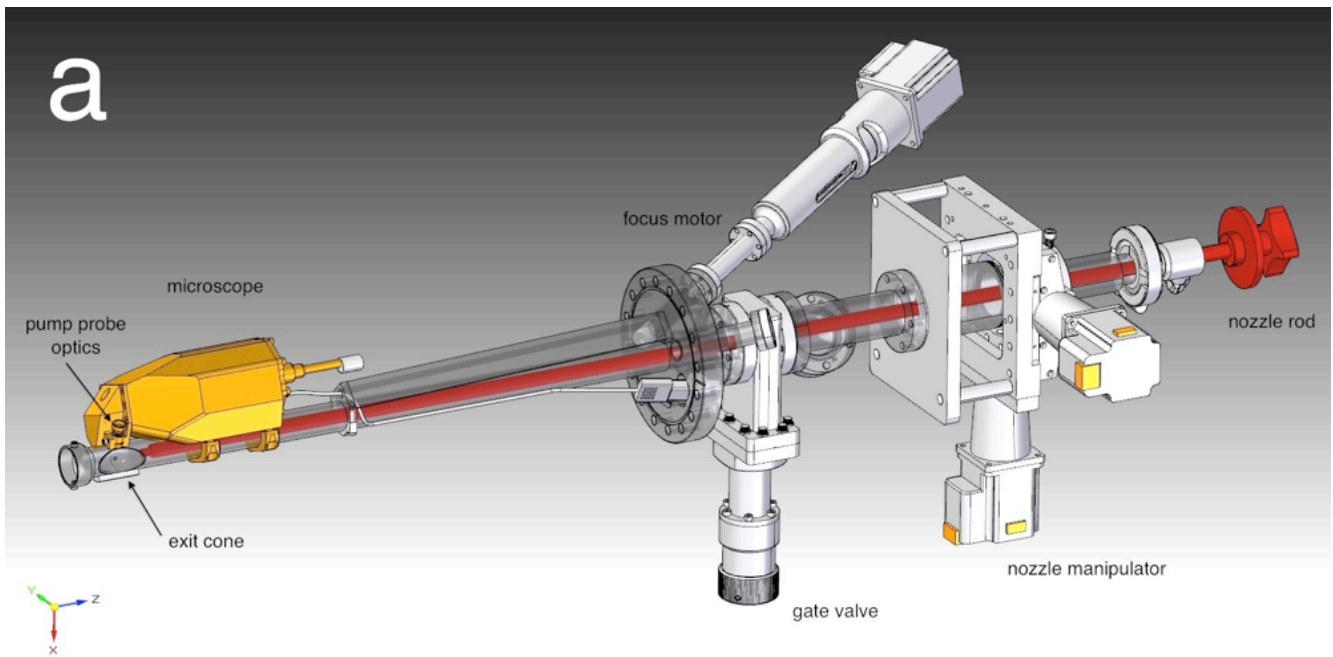

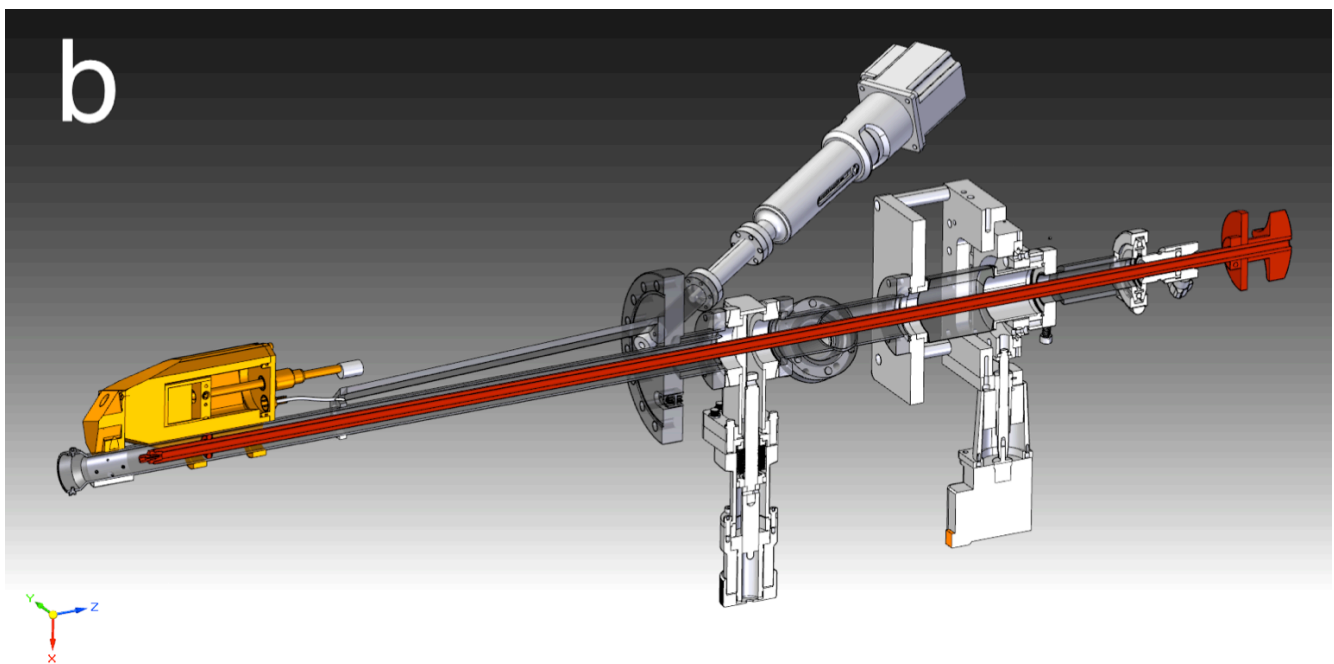

Figure 9: Nozzle shroud (a) and cross section of nozzle shroud (b). Some parts are shown transparent. The nozzle rod, which holds the nozzle at its end, is shown in red.

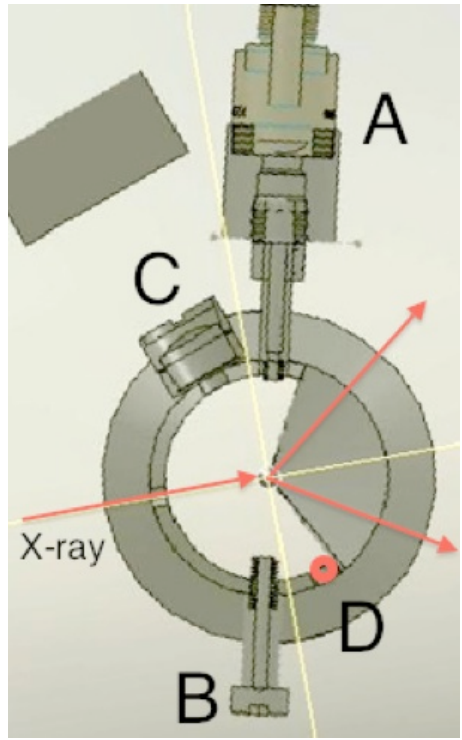

Figure 10: Cross section through differential pumping shroud at the interaction region. The X-ray beam enters from the left through a 2mm diameter hole and hits the liquid stream in the center (stream direction out of the paper plane). Diffracted X-rays leave the shroud through the exit cone on the right, with a maximum angle of 60 degrees. The pump laser optics (A), pump laser beam dump (B) and microscope objective lens (C) and LED (D) are also shown.

### c. Microscope and Pump Laser

During previous beamtimes at the LCLS and at FLASH, the liquid nozzle has always been operated without direct visual feedback. The feedback was indirect in the sense that the presence of a diffraction streak was the indication for the existence of a liquid jet. This diagnosis of the jet condition is inaccurate and incomplete, i.e. it cannot be determined whether the nozzle freezes in vacuum or the jet changes direction due to accumulation of sample debris at the nozzle end. Adjusting the gas and liquid pressures to optimize jet conditions is much easier when direct optical feedback is provided. Therefore an in vacuum microscope was developed which allows direct observation of the jet from the control room. The microscope has to be small, vacuum compatible and one has to be able to change the focus from outside the vacuum.

The jet is best visible when back illuminated, therefore a LED had to be mounted in the shroud opposite to the microscope objective. The microscope is attached to the nozzle shroud (Figure 12 and Figure 11). To make the microscope as compact as possible, its optical axis is bend by 90 degree with a mirror. A single lens is used to project a magnified image of the nozzle onto a CCD chip (Moticam 2300, 3 Megapixels). The whole microscope is enclosed in a vacuum tight container at ambient pressure and is sealed against the chamber vacuum. Image focusing is achieved from outside the chamber by moving the CCD camera along the optic axis with a motorized push rod inside a flexible bellows. Since there is only one lens, focusing changes also the magnification. The microscope field of view is about 600 x 800 micron. The CCD camera electronics is connected to the outside by an in vacuum USB cable. Figure 13 shows an image of the liquid jet collected with this microscope at an X-ray energy of 1.8kV. The area where the LCLS X-ray beam hit the jet is visible as a bright dot, which is caused by the plasma that is created by the high intensity X-ray pulses.

The microscope housing also contains a fiber optic coupling and lens to focus a pump laser onto the jet. This allows pump probe experiments where the sample (biomolecule in the jet) is excited with an optical laser and after a preset time delay probed with the X-ray pulse. The maximum time delay between pump and probe is determined by the size of the laser spot, the position of the (much smaller) X-ray spot on the jet and the jet speed. The laser is coupled into vacuum via a fiber optic feedthrough (Accu-Glass) with a 100 micron diameter multimode fiber. The end of the fiber is imaged one to one with a lens onto the jet, i.e. the laser spot size on the jet is 100 micron. Changing the position of the lens allows to increase the laser spot size. Figure 14 shows an image taken with the in vacuum microscope while the pump laser (YAG fiber laser (JEDI), 532nm, 6nsec pulse duration with 10 microsecond pump-probe time delay) spot was aligned with the jet.

This injector system has been used at the LCLS in June 2010 and February 2011. During these experiments the jet has been in operation continuously for a week while changing to several different liquid samples. Typical flow rates where 10 - 20 microliter/minute and the jet diameter was about 6 micrometer. While the liquid jet was running, the vacuum in the experimental chamber (CAMP and CXI) remained at $< 3 \times 10^{-5}$ Torr. During nozzle change the chamber vacuum stayed at $<1 \times 10^{-6}$ Torr.

The liquid sample supply system uses a Rheodyne® MX Series II™ automated fluidic valve, which allows switching between two liquids contained in HPLC sample loops. The liquid and gas pressure is supplied by remote controlled high-pressure gas regulators. Alternatively the liquid pressure can also be applied with a HPLC pump. This system allows uninterrupted flow to the nozzle even when changing samples and thereby completely eliminates the turn-off/turn-on transients that can disrupt the liquid jet. It also provides the means to fill or rinse a nozzle supply reservoir such that the liquid contains no air bubbles, which can interrupt the flow and cause icing at the nozzle. Figure 15 shows a diffraction pattern obtained from a single nanocrystal of Photosystem I in the liquid jet at the LCLS with an X-ray pulse length of 70fsec, $10^{12}$ photons/pulse, 6.9 Å wavelength, 8.5 Å resolution at the corner of the detector. For a detailed account of these experiments see [5].

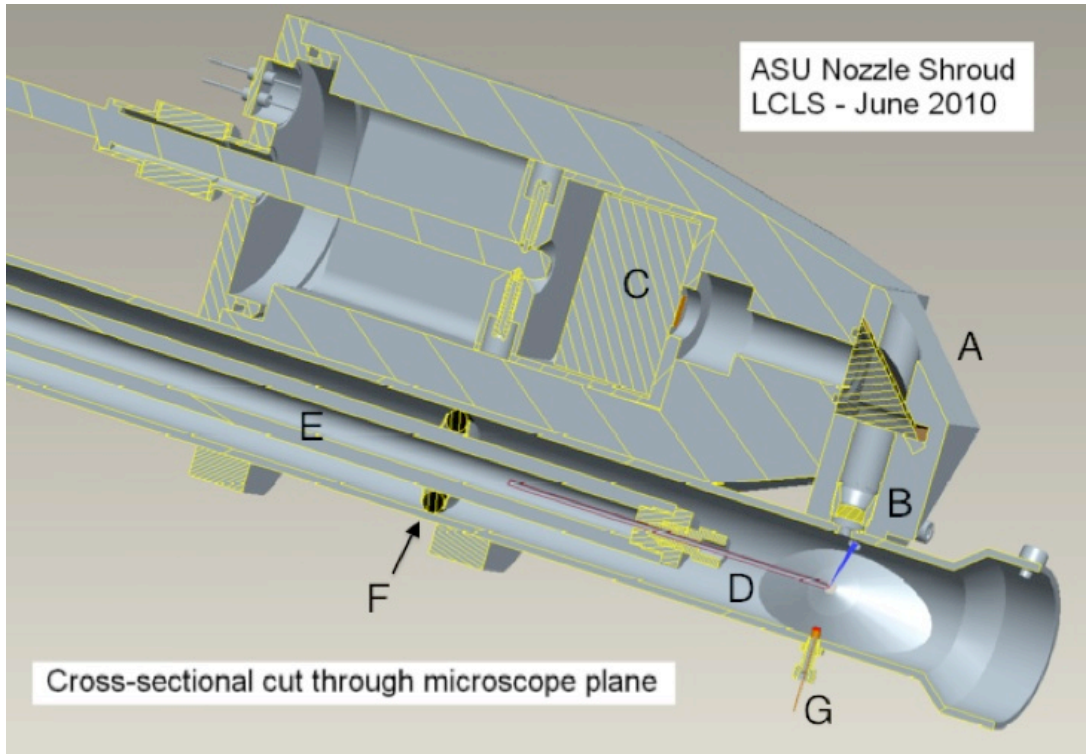

Figure 11: Cross section through in-vacuum microscope attached to differential pumping shroud. The CCD chip with electronics and the optics is contained in a vacuum sealed housing which is at ambient pressure. This avoids overheating of the CCD electronics. The optics consists of a mirror (A) and a single lens (B). The distance between CCD module (C) and mirror can be adjusted by a push rod to focus the microscope. Also visible are the nozzle (D), the nozzle rod (E) with viton gasket (F) and the LED (G). The pump laser optics is behind the mirror and is not visible (pump laser beam is shown in blue.

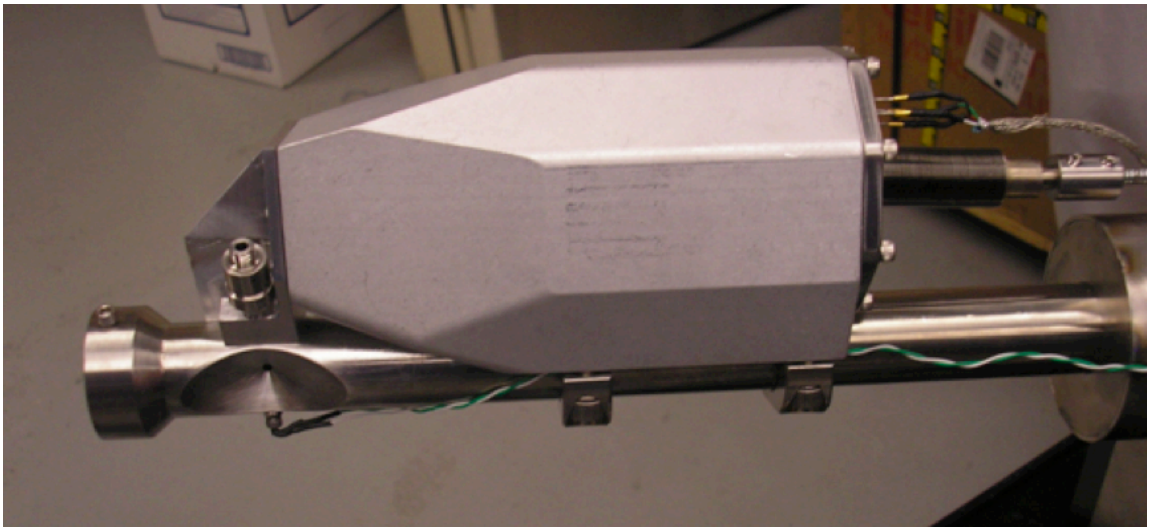

Figure 12: In vacuum microscope attached to nozzle shroud.

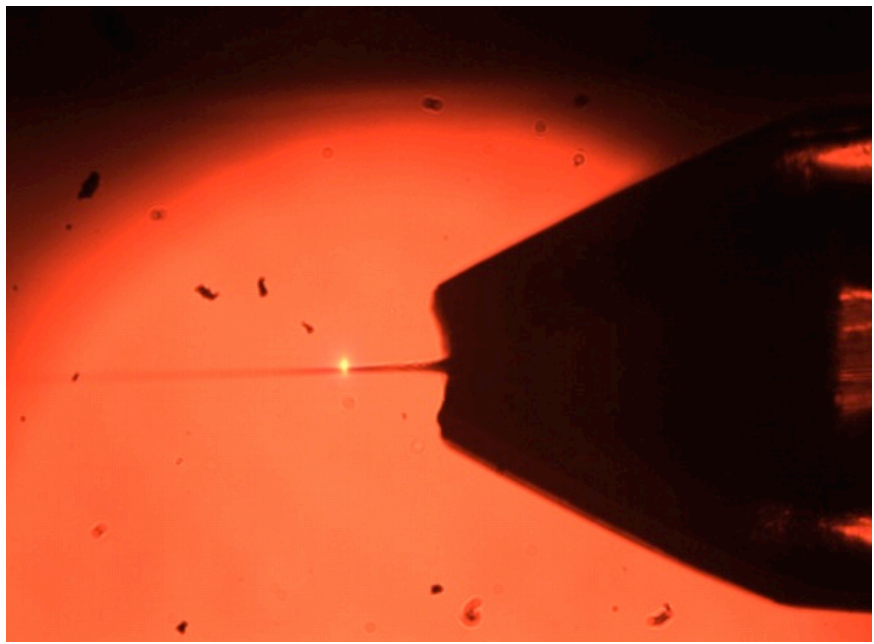

Figure 13: Nozzle and liquid jet imaged with in vacuum microscope, while LCLS X-ray beam hits the jet. The spot where the X-ray beam hits the jet is visible as a bright area due to the plasma created by the high intensity X-ray pulses. The repeat frequency of the X-ray pulses was 60Hz. Dark specks are dust on the CCD glass cover.

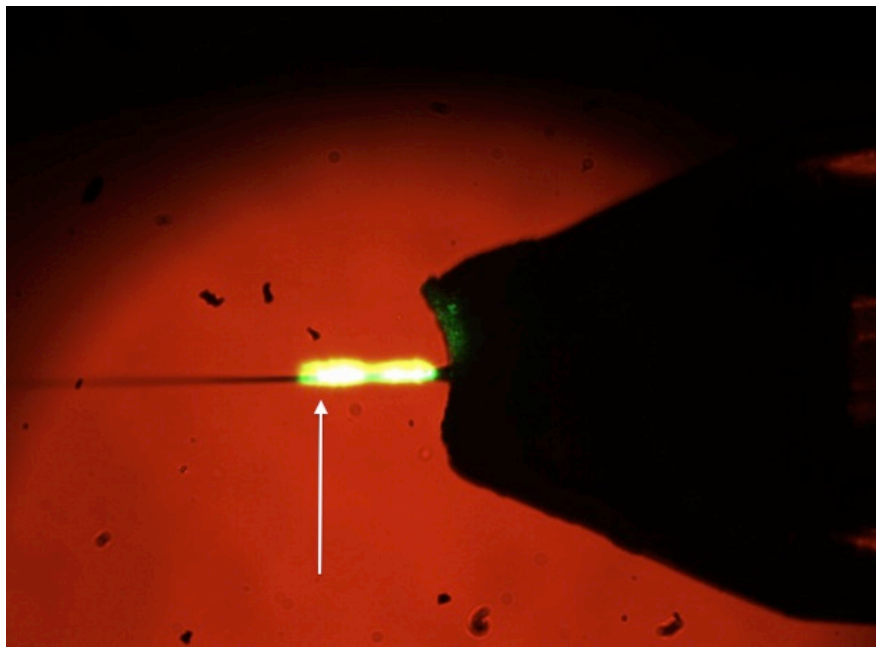

Figure 14: Pump probe experiments with Photosystem II: Pump laser aligned with the liquid jet. The image was recorded with the in-vacuum microscope. The nozzle is visible to the right, the jet (containing PS II nanocrystals) is visible exiting the nozzle. The bright green area is the pump laser spot intersecting the jet. The arrow shows the position of the XFEL beam. The time delay between the pump laser and the X-ray pulse was 10 microseconds.

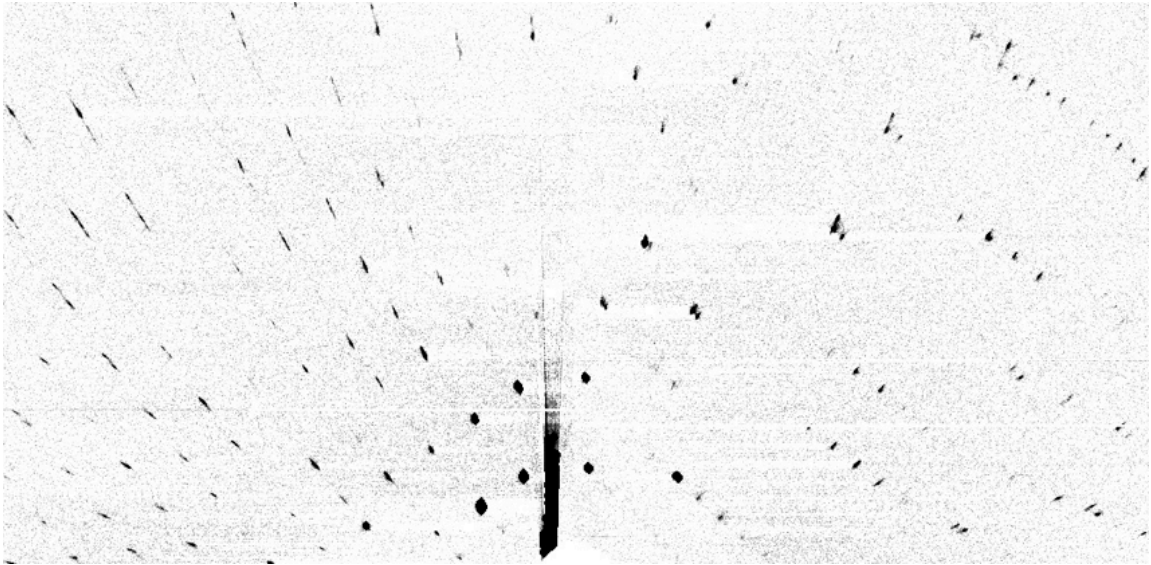

Figure 15: Femtosecond diffraction pattern from single Photosystem I nanocrystal in the liquid jet showing Bragg diffraction out to 0.85 nm resolution (wavelength 0.69nm, pulse duration 70fsec, $10^{12}$ photons/pulse).

**Conclusions:**
A liquid jet injector system has been developed that can be used for hydrated sample delivery at X-ray Free Electron Laser (XFEL) sources and 3rd generation synchrotron sources. The use of a liquid jet injector is preferable where complete sample hydration is mandatory and high hit rate is required. The liquid injector has been used during several experimental runs at the LCLS and has allowed the collection of high quality femtosecond diffraction patterns from many different protein nanocrystals. The injector design avoids contamination of the experimental chamber and CCD detector by the inherently vacuum incompatible biomolecular buffer solutions that are sprayed into X-ray beam. Furthermore the nozzle design avoids clogging problems that plague all injector designs using a simple Rayleigh jet. The use of GDVN allows also much lower flow rates than attainable with conventional Rayleigh nozzles. All motions of the injector are motorized and controllable via EPICs software. The incorporated pump laser optics allows pump probe experiments where biomolecules are excited by a laser pulse to cause conformational changes, which are then probed by the following X-ray pulse. First pump probe experiments on Photosystem II have been carried out and are currently being analyzed. Due to the current X-ray pulse repetition rate of 60Hz at the LCLS, there is still a large amount of sample wasted between pulses, but new XFEL sources (NGLS and European XFEL) will have higher repetition rates and will thereby minimize sample consumption. Another possibility, which is currently being examined is to reduce the liquid flow rate of the GDVN system.

**Acknowledgements:**
We gratefully acknowledge support by the US National Science Foundation (award MCB 0919195). Experiments with the injector were carried out at the Linac Coherent Light Source and the Advanced Light Source, both National User Facilities operated respectively by Stanford University and the University of California on behalf of the US Department of Energy (DOE), Office of Basic Energy Sciences. We acknowledge support from the DOE through the PULSE Institute at the SLAC National Accelerator Laboratory.